\PassOptionsToPackage{unicode}{hyperref}
\PassOptionsToPackage{hyphens}{url}
\PassOptionsToPackage{dvipsnames,svgnames,x11names}{xcolor}
\documentclass[
]{article}

\usepackage{amsmath,amssymb}
\usepackage{iftex}
\ifPDFTeX
  \usepackage[T1]{fontenc}
  \usepackage[utf8]{inputenc}
  \usepackage{textcomp} 
\else 
  \usepackage{unicode-math}
  \defaultfontfeatures{Scale=MatchLowercase}
  \defaultfontfeatures[\rmfamily]{Ligatures=TeX,Scale=1}
\fi
\usepackage{lmodern}
\ifPDFTeX\else  
\fi
\IfFileExists{upquote.sty}{\usepackage{upquote}}{}
\IfFileExists{microtype.sty}{
  \usepackage[]{microtype}
  \UseMicrotypeSet[protrusion]{basicmath} 
}{}
\makeatletter
\@ifundefined{KOMAClassName}{
  \IfFileExists{parskip.sty}{%
    \usepackage{parskip}
  }{
    \setlength{\parindent}{0pt}
    \setlength{\parskip}{6pt plus 2pt minus 1pt}}
}{
  \KOMAoptions{parskip=half}}
\makeatother
\usepackage{xcolor}
\ifx\paragraph\undefined\else
  \let\oldparagraph\paragraph
  \renewcommand{\paragraph}[1]{\oldparagraph{#1}\mbox{}}
\fi
\ifx\subparagraph\undefined\else
  \let\oldsubparagraph\subparagraph
  \renewcommand{\subparagraph}[1]{\oldsubparagraph{#1}\mbox{}}
\fi

\usepackage{longtable,booktabs,array}
\usepackage{calc} 
\usepackage{etoolbox}
\makeatletter
\patchcmd\longtable{\par}{\if@noskipsec\mbox{}\fi\par}{}{}
\makeatother
\IfFileExists{footnotehyper.sty}{\usepackage{footnotehyper}}{\usepackage{footnote}}
\makesavenoteenv{longtable}
\usepackage{graphicx}
\makeatletter
\def\maxwidth{\ifdim\Gin@nat@width>\linewidth\linewidth\else\Gin@nat@width\fi}
\def\maxheight{\ifdim\Gin@nat@height>\textheight\textheight\else\Gin@nat@height\fi}
\makeatother
\setkeys{Gin}{width=\maxwidth,height=\maxheight,keepaspectratio}
\makeatletter
\def\fps@figure{htbp}
\makeatother
\NewDocumentCommand\citeproctext{}{}

\makeatletter
 \let\@cite@ofmt\@firstofone
 \def\@biblabel#1{}
 \def\@cite#1#2{{#1\if@tempswa , #2\fi}}
\makeatother
\newlength{\cslhangindent}
\newlength{\csllabelwidth}
\newenvironment{CSLReferences}[2] 
 {\begin{list}{}{%
  \setlength{\itemindent}{0pt}
  \setlength{\leftmargin}{0pt}
  \setlength{\parsep}{0pt}
  \ifodd #1
   \setlength{\leftmargin}{\cslhangindent}
   \setlength{\itemindent}{-1\cslhangindent}
  \fi
  \setlength{\itemsep}{#2\baselineskip}}}
 {\end{list}}
\usepackage{calc}

\usepackage{multirow}
\usepackage{arxiv}
\usepackage{orcidlink}
\usepackage{amsmath}
\usepackage[T1]{fontenc}
\makeatletter
\@ifpackageloaded{caption}{}{\usepackage{caption}}
\AtBeginDocument{%
\ifdefined\contentsname
  \renewcommand*\contentsname{Table of contents}
\else
  \newcommand\contentsname{Table of contents}
\fi
\ifdefined\listfigurename
  \renewcommand*\listfigurename{List of Figures}
\else
  \newcommand\listfigurename{List of Figures}
\fi
\ifdefined\listtablename
  \renewcommand*\listtablename{List of Tables}
\else
  \newcommand\listtablename{List of Tables}
\fi
\ifdefined\figurename
  \renewcommand*\figurename{Figure}
\else
  \newcommand\figurename{Figure}
\fi
\ifdefined\tablename
  \renewcommand*\tablename{Table}
\else
  \newcommand\tablename{Table}
\fi
}
\@ifpackageloaded{float}{}{\usepackage{float}}
\floatstyle{ruled}
\@ifundefined{c@chapter}{\newfloat{codelisting}{h}{lop}}{\newfloat{codelisting}{h}{lop}[chapter]}
\floatname{codelisting}{Listing}

\makeatother
\makeatletter
\@ifpackageloaded{caption}{}{\usepackage{caption}}
\@ifpackageloaded{subcaption}{}{\usepackage{subcaption}}
\makeatother
\ifLuaTeX
  \usepackage{selnolig}  
\fi
\IfFileExists{bookmark.sty}{\usepackage{bookmark}}{\usepackage{hyperref}}
\IfFileExists{xurl.sty}{\usepackage{xurl}}{} 
\hypersetup{
  pdftitle={Overture POI data for the United Kingdom: a comprehensive, queryable open data product},
  pdfauthor={Patrick Ballantyne; Cillian Berragan},
  pdfkeywords={Points of Interest, Overture, Amazon Web Services},
  colorlinks=true,
  linkcolor={blue},
  filecolor={Maroon},
  citecolor={Blue},
  urlcolor={Blue},
  pdfcreator={LaTeX via pandoc}}

\newcommand{\runninghead}{A Preprint }
\title{Overture POI data for the United Kingdom: a comprehensive,
queryable open data product}
\def\asep{\\\\\\ } 
\def\asep{\And }
\author{\textbf{Patrick
Ballantyne}~\orcidlink{0000-0001-8980-2912}\\Geographic Data Science
Lab\\University of
Liverpool\\\\\href{mailto:p.ballantyne@liverpool.ac.uk}{p.ballantyne@liverpool.ac.uk}\asep\textbf{Cillian
Berragan}~\orcidlink{0000-0003-2198-2245}\\Geographic Data Science
Lab\\University of
Liverpool\\\\\href{mailto:c.berragan@liverpool.ac.uk}{c.berragan@liverpool.ac.uk}}
\date{}
\begin{document}
\maketitle
\begin{abstract}
Point of Interest data that is comprehensive, globally-available and
open-access, is sparse, despite being important inputs for research in a
number of application areas. New data from the Overture Maps Foundation
offers significant potential in this arena, but accessing the data
relies on computational resources beyond the skillset and capacity of
the average researcher. In this article, we provide a processed version
of the Overture places (POI) dataset for the UK, in a fully-queryable
format, and provide accompanying code through which to explore the data,
and generate other national subsets. In the article, we describe the
construction and characteristics of the dataset, before considering how
reliable it is (locational accuracy, attribute comprehensiveness),
through direct comparison with Geolytix supermarket data. This dataset
can support new and important research projects in a variety of
different thematic areas, and foster a network of researchers to further
evaluate its advantages and limitations.
\end{abstract}
{\bfseries \emph Keywords}
\def\sep{\textbullet\ }
Points of Interest \sep Overture \sep 
Amazon Web Services

\section{Introduction}\label{introduction}

Point of Interest (POI) data is an invaluable source of information,
acting as a key input to much of the research that has, and continues to
be generated in urban analytics and city science. These data provide key
locational attributes about a broad variety of social, environmental and
economic phenomena, including historical landmarks, parks, hospitals and
retailers, and have been vital sources of data for different
applications, including health (Green et al. 2018; Hobbs et al. 2019),
urban mobility (Graells-Garrido et al. 2021; Jay et al. 2022), retail
and location analysis (Ballantyne et al. 2022), transportation (Owen,
Arribas-Bel, and Rowe 2023; Credit 2018), and many others. However, a
major challenge when working with POI data relates to the coverage and
comprehensiveness of these datasets (Ballantyne et al. 2022; Zhang and
Pfoser 2019). By this we mean how much the chosen source(s) of POI data
restricts the analyses to specific cities or regions (i.e., coverage),
and the attributes and characteristics that are provided for each POI
(i.e., comprehensiveness).

Many POI datasets offer a high level of global coverage and
availability, such as OpenStreetMap. However there are problems when
considering the coverage and comprehensiveness of OpenStreetMap data at
finer spatial resolutions and in areas with less contributors (Haklay
2010), as well as in less developed countries (Mahabir et al. 2017).
Similarly, datasets like OpenStreetMap often contain inconsistent
attributes for economic activities like retail stores and leisure (Zhang
and Pfoser 2019; Ballantyne et al. 2022). Some POI datasets exist which
fill this gap, such as the Ordnance Survey `Points of Interest' data
product, which provides a more comprehensive database of economic
activities (Haklay 2010 ), but is not openly-available. Other data
providers have democratised access to comprehensive POI datasets such as
SafeGraph and the Local Data Company, however these datasets exhibit
poor global coverage of non-branded POIs (SafeGraph), and a lack of
comprehensive coverage in the UK (Dolega et al. 2021). As a result,
there is a clear gap for data that can address some of these
limitations, by providing an openly-available, comprehensive and
accurate source of POIs for the UK. In this article, we introduce
readers to a processed version of the Overture Maps places (POI) dataset
(Overture Maps Foundation 2023), which arguably provides a strong
solution to many of these problems, and can facilitate groundbreaking
urban analytics research in a number of different application areas.

\section{Data}\label{data}

The data was accessed through the Overture Maps Foundation, which was
set up as a collaborative venture to develop reliable, easy-to-use, and
interoperable open map data (Overture Maps Foundation 2023). The
foundation, which is steered by Amazon, Microsoft, Meta and TomTom, has
developed a number of open data products including Buildings, Places,
Transportation and Administrative Geographies, all of which are
available at global scales and contain a detailed number of attributes
(Overture Maps Foundation 2023). Users can access the data parquet files
from the cloud using Amazon Athena, Microsoft Synapse or DuckDB, or
download them locally. However, a specific challenge for urban analytics
researchers and city scientists is that the majority will not have the
data engineering skills to query these datasets from the cloud, and
process the attributes in their nested JSON format. Furthermore, for
those who want to download the files locally, they can be difficult to
work with, as the full global places file is over 200 GB. Therefore, our
aim is to provide a processed subset of the Overture places dataset for
the UK, which bypasses these issues, and creates an open data product
for use in research.

Overture hosts all data through Amazon Web Services (AWS), which enables
a number of query end points to be used to download data subsets. The
Overture data schema includes a bounding box structure column to enable
efficient spatial SQL queries. To query POI data for the UK, a spatial
SQL query was constructed using the DuckDB SQL engine and the UK
bounding box, based on EPSG:27700. This query downloaded a GeoPackage
file containing all POIs within the UK bounding box, totalling 1.34 GB.
This file was then clipped to the administrative boundaries of the
United Kingdom, to exclude non-UK places that appeared within the
bounding box query.

As noted, many of the columns that provide metadata relating to POIs are
represented in a nested JSON format (columns containing lists of lists),
which are difficult to efficiently parse with traditional tabular data
frame libraries. We therefore processed the following columns to ensure
the data frame remained two-dimensional: Names, Category, Address and
Brand. Following this processing, we spatially joined the 2021 census
area geographies for England including Output Areas (OA), Lower layer
Super Output Areas (LSOA), Middle layer Super Output Areas (MSOA), and
the 2022 Local Authority Districts (LAD). For both Scotland and Northern
Ireland, we spatially joined the 2011 Data Zone geographies. We also
include the H3 (hexagons) addresses associated with each point for all
resolutions between 1 and 9. The resulting dataset is a 358 MB
GeoParquet file, hosted as part of a DagsHub data repository, and the
final processed data file, comprising the Overture POI subset for the UK
can be easily downloaded\footnote{https://figshare.com/s/144265a705159c03c08f?file=42761512}.
A list of attributes for the data product can be found in Table i
(supplementary materials), and as a secondary output of this paper, an
example workflow for how to extract Overture places for other study
areas has also been produced\footnote{https://figshare.com/s/144265a705159c03c08f?file=42809656}.

\section{Reliability Analysis - Retail
Brands}\label{reliability-analysis---retail-brands}

To assess the reliability of Overture places, we compared them with the
Geolytix Supermarket Retail Points dataset (Geolytix 2023), which is
known to provide reliable information about supermarkets in the UK, and
provides a useful `ground-truth' dataset to test how well Overture
represents economic activities. In particular, we examined how many of
the Geolytix supermarkets are captured in Overture, the accuracy of the
POI coordinates, and how complete the category/brand information is.
Table \ref{table1} shows that the Overture data aligns well with the
Geolytix data, with small differences across the three retailers
(\textless{} 5\%). Table \ref{table1} shows that there was a relatively
low median distance (metres) between Overture points and their closest
Geolytix point, evidencing a relatively high level of accuracy in terms
of geographical positioning. This is an important attribute, as
incorrect positioning of POI data can have dramatic implications for
accessibility measurement (Green et al. 2018; Graells-Garrido et al.
2021), and urban boundary delineation (Ballantyne et al. 2022).

In terms of the comprehensiveness of the category and brand information,
a large number of the Overture POIs contained missing values for
categories or brands (Table \ref{table2}), making filtering of the
dataset to a specific retailer (e.g., Waitrose), slightly less simple.
Table \ref{table2} displays the complexity of these issues, where
different degrees of completeness are apparent when considering the
source of the POI (Meta or Microsoft). This has strong implications for
how Overture data can and should be used, especially for applications
involving specific POI categories or brands. Whilst it is not impossible
to extract a complete list of POIs for a retailer, through collective
filtering of POI name, brand and categories to collect these features
(see supplementary materials), users should be aware of the high level
of attribute incompleteness for POIs extracted from Microsoft. Further
reliability analysis is beyond the scope of this paper, but there is a
clear need for further investigation into how well Overture places
captures category and brand information for other non-retail POIs (e.g.,
GP practices, post offices).

\def\arraystretch{1.5}
\begin{table}[]
\caption{\label{table1} Reliability analysis of Overture compared with Geolytix retail points dataset.}
\centering
\bigskip
\begin{tabular}{lccc}
\hline
\textbf{Retailer} & \textbf{Geolytix count} & \textbf{Overture count} & \textbf{Average distance between points (m)} \\
\hline
Waitrose & 422            & 420            & 8.3                                 \\
Spar     & 2,339          & 2,308          & 6.5                                 \\
Tesco    & 2,840          & 2,753          & 6.2 \\
\hline
\end{tabular}
\end{table}

\def\arraystretch{1.5}
\begin{table}[]
\caption{\label{table2} Overture attributes compared with Geolytix retail points dataset.}
\centering
\bigskip
\begin{tabular}{lcccc}
\hline
\multirow{2}{*}{} & \multicolumn{4}{c}{\textbf{Attribute incompleteness (\%)}}                                                  \\ \cline{2-5} 
                  & \multicolumn{2}{c|}{\textbf{Category information}}                  & \multicolumn{2}{c}{\textbf{Brand information}} \\ \hline
\textbf{Retailer}          & \multicolumn{1}{|c}{\textbf{Meta}} & \multicolumn{1}{c|}{\textbf{Microsoft}} & \multicolumn{1}{c}{\textbf{Meta}}  & \textbf{Microsoft} \\ \hline
Waitrose          & \multicolumn{1}{|c}{100}  & \multicolumn{1}{c|}{N/A}       & \multicolumn{1}{c}{23.33} & N/A       \\
Spar              & \multicolumn{1}{|c}{0.18} & \multicolumn{1}{c|}{100.00}    & \multicolumn{1}{c}{11.63} & 100.00    \\
Tesco             & \multicolumn{1}{|c}{0.00} & \multicolumn{1}{c|}{N/A}       & \multicolumn{1}{c}{2.29}  & N/A       \\ \hline
\end{tabular}
\end{table}

\section{Application - Mapping supermarkets in the
UK}\label{application---mapping-supermarkets-in-the-uk}

To demonstrate how this dataset can be used, an example workflow has
been presented which reads in the UK processed version of Overture
places, filters to a specific brand of supermarket, and then maps the
distribution of these nationally (Figure 1). The purpose of these
workflows is to illustrate how easy it is to work with this dataset, and
the variety of different POI attributes that are stored within the
dataset. Example workflows have been presented for both the
Python\footnote{https://figshare.com/s/144265a705159c03c08f?file=42809500}
and R\footnote{https://figshare.com/s/144265a705159c03c08f?file=42809452}
programming languages, and utilise preferred packages for data
manipulation and mapping (e.g., arrow, geopandas)

\begin{figure}

\centering{

\captionsetup{labelsep=none}\includegraphics{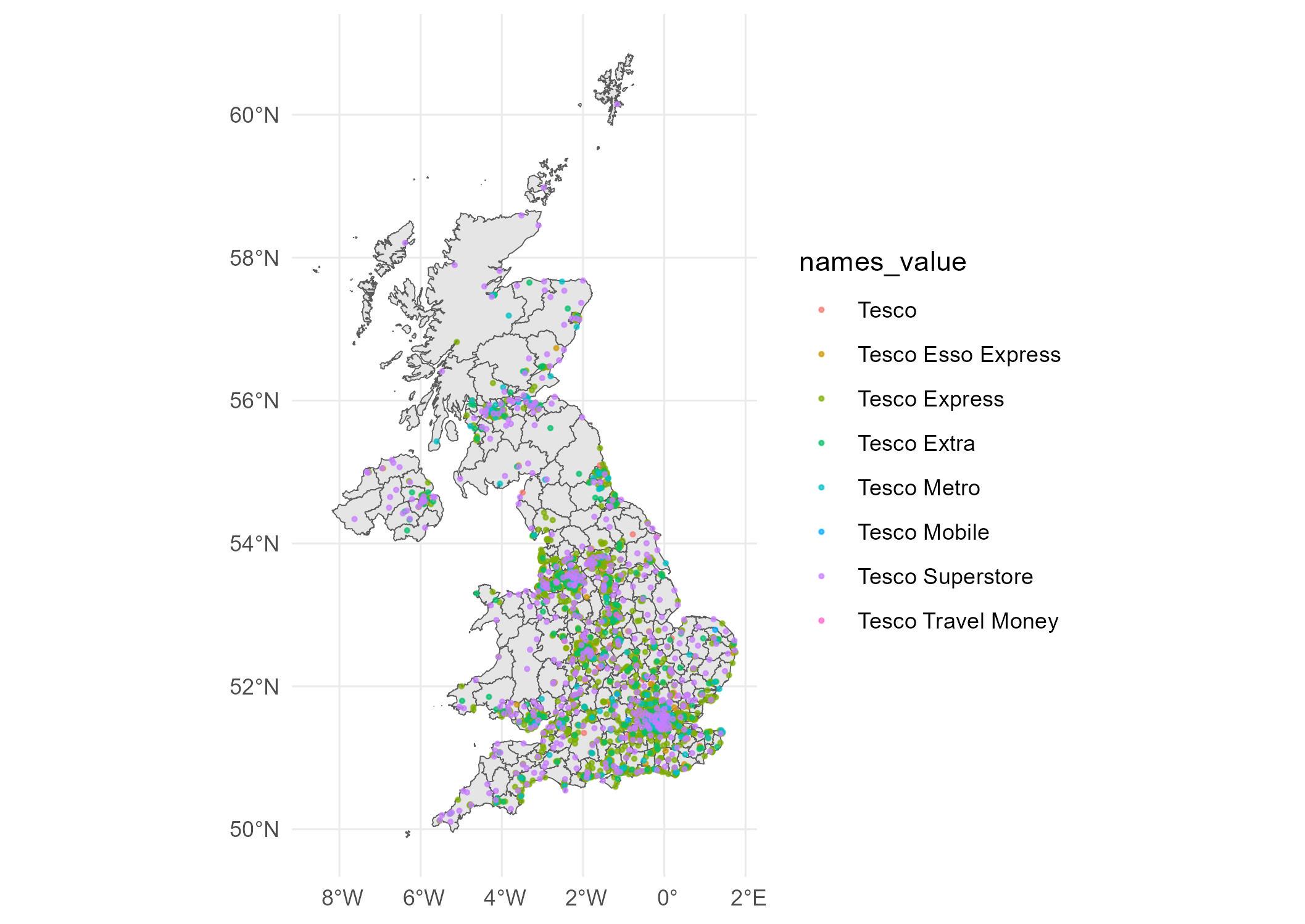}

}

\caption{\label{fig-figure1}}

\end{figure}%

\section{Conclusion}\label{conclusion}

This paper presents a comprehensive, queryable open data product, which
represents a processed UK national subset of the Overture places
database. This new open data product makes Overture data more accessible
for researchers, bypassing the need for advanced data engineering skills
and large amounts of memory on which to store the complete database. The
potential applications of this data product in a variety of different
fields is highly significant (e.g., urban accessibility), given the
evidence presented about the coverage, comprehensiveness and locational
accuracy of this dataset. At a time where the retail sector is
undergoing significant transformations in response to the cost-of-living
crisis, such data can provide invaluable insights about the
characteristics and performance of the sector (Ballantyne et al. 2022;
Dolega et al. 2021), which has historically been a challenge due to the
availability of suitable retailer data. However, there are inherent
limitations to this dataset, which have been illustrated through direct
comparison with Geolytix data. Users need to be cautious about how they
are using this data, especially when the POIs they are using are largely
sourced from Microsoft. However, it is our hope that by releasing this
data into the open domain, a network of researchers will be fostered who
can utilise this data for their own research questions, and critically
evaluate how the Overture places database represents a variety of
different social, economic and environmental activities.

\section{Data Availability Statement}\label{data-availability-statement}

The UK Overture data product (anonymised for peer review) can be
downloaded directly from Figshare:
\url{https://figshare.com/s/144265a705159c03c08f?file=42761512}. The
data product can be directly queried from the DagsHub repository, but
for the purposes of anonymous peer review, this has not been included in
the paper.

\section*{References}\label{references}
\addcontentsline{toc}{section}{References}

\phantomsection\label{refs}

\end{document}


\section*{Supplementary material}\label{supplementary-material}
\addcontentsline{toc}{section}{Supplementary material}

\renewcommand{\thetable}{\roman{table}}
\renewcommand{\thefigure}{\roman{figure}}

\emph{Overture data product attributes}

\def\arraystretch{1.5}
\begin{table}[H]
\caption{\label{tablei} Attributes in Overture UK open data product, with an additional indicator describing which attributes were processed out of their nested JSON format.}
\centering
\bigskip
\begin{tabular}{p{6cm}p{6cm}c}
\hline
\multicolumn{1}{c}{\textbf{Attribute}}                                                                        & \multicolumn{1}{c}{\textbf{Description}}                                                                     & \textbf{Processed} \\
\hline
id                                                                                                   & Unique ID assigned by Overture.                                                                     & N         \\
updatetime, version                                                                                  & Information about the version (and date) of the Overture data - most recent.                        & N         \\
confidence                                                                                           & Attribute assigned by Overture                                                                      & N         \\
websites, socials, emails, phones                                                                    & Contact information and social media associated with POI.                                           & N         \\
names\_value                                                                                         & POI name.                                                                                           & Y         \\
category\_main, category\_alternative                                                                & Assigned category for POI.                                                                          & Y         \\
addresses\_postcode, addresses\_freeform, addresses\_country, addresses\_locality, addresses\_region & Address information for POI.                                                                        & Y         \\
sources\_dataset                                                                                     & Data source (meta or microsoft)                                                                     & N         \\
lat, lng                                                                                             & Coordinates for POI.                                                                                & N         \\
h3\_01 ... h3\_09                                                                                    & H3 addresses for each POI.                                                                          & N         \\
easting, northing                                                                                    & Coordiantes for POI.                                                                                & N         \\
LAD22CD ... LGD2014\_nm                                                                              & Administrative geographies for each POI, different for England/Wales, Scotland and Northern Ireland & N         \\
\hline
\end{tabular}
\end{table}

\emph{Scraping Overture data for specific brands}

As discussed in the main body, extracting a complete list of POIs for a
brand (e.g., Waitrose) is difficult due to missing values in the
category and brand columns. For example, selecting all POIs by brand, as
below in Figure i, results in 0 POIs for Waitrose, and 1,934 POIs for
Spar.

\begin{figure}[H]

{\centering \includegraphics{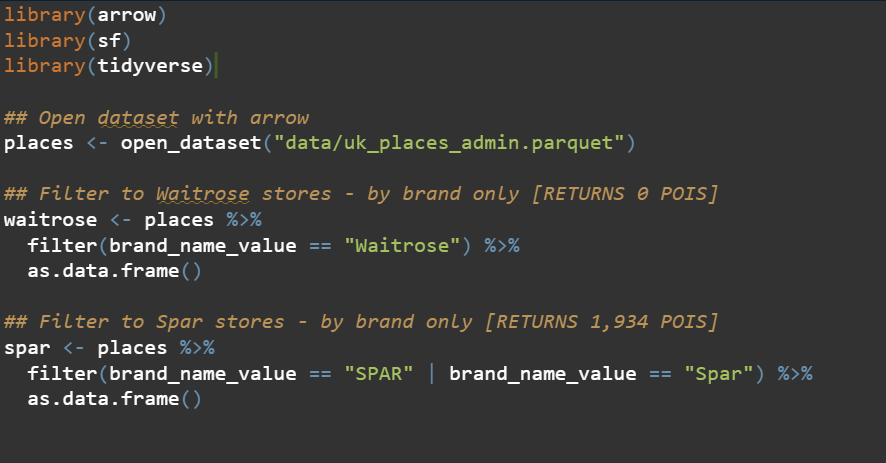}

}

\caption{R code snippet highlighting brand filtering of POIs.}

\end{figure}%

However, by integrating additional filters, specifically on the POI name
and the POI category columns, you can get to the final list of POIs for
each retail brand, which closely resemble those in the Geolytix dataset.
For example, in Figure ii we show that the brand and name columns need
to be filtered to extract all Spar supermarket stores, and name and
category columns need to be filtered to extract all Waitrose
supermarkets (excluding petrol stations and other retail formats). This
code snippet (Figure ii) results in 420 POIs for Waitrose, and 2,308
POIs for Spar, as in Table 1. This has strong implications for future
use of this dataset, where users should consider POI name, brand and
category if seeking to extract a complete list of POIs for specific
research questions or applications.

\begin{figure}[H]

{\centering \includegraphics{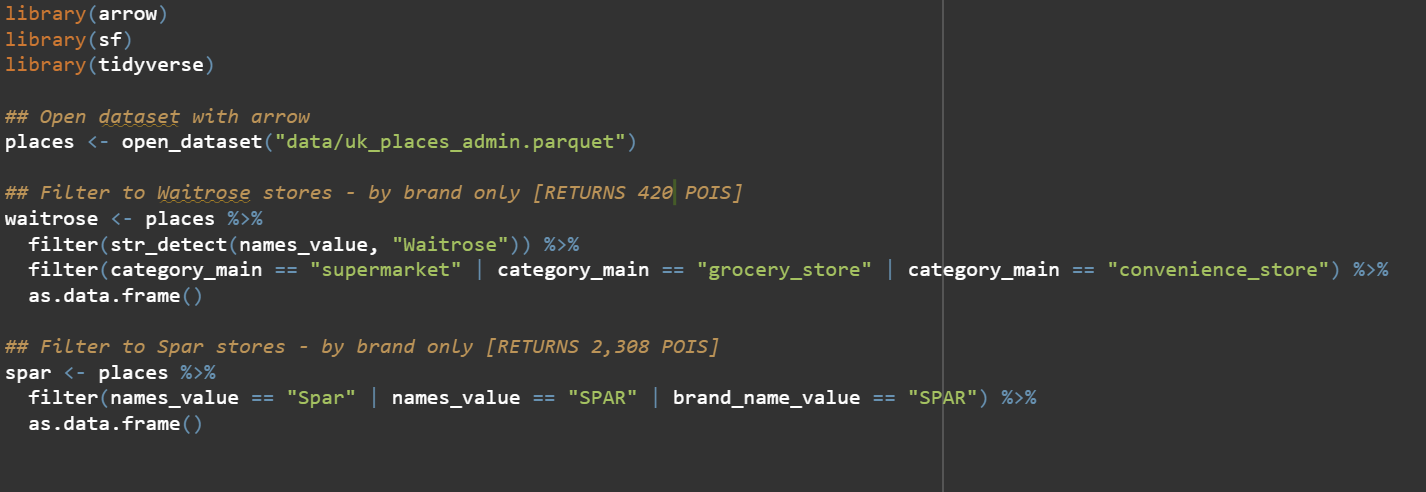}

}

\caption{Code snippet highlighting additional filtering of POI
attributes.}

\end{figure}%